\documentclass[superscriptaddress,twocolumn,showpacs,pre]{revtex4}

\usepackage{graphicx}  
\usepackage{bm}
\usepackage{times}
\usepackage{amsmath}
\newcommand{\bra}[1]{\left(#1\right)}
\newcommand{\Bra}[1]{\left[#1\right]}

\begin{document}

\title{Front motion and localized states in an asymmetric bistable activator-inhibitor system with saturation}

\author{Arik Yochelis}
\affiliation{Department of Medicine (Cardiology), University of California, Los Angeles, CA 90095}
\author{Alan Garfinkel}
\affiliation{Department of Medicine (Cardiology), University of California, Los Angeles, CA 90095} \affiliation{Department of Physiological
Science, University of California, Los Angeles, CA 90095}

\received{\today}
\begin{abstract}
We study the spatiotemporal properties of coherent states (peaks, holes, and fronts) in a bistable activator-inhibitor system that exhibits
biochemical saturated autocatalysis, and in which fronts do not preserve spatial parity symmetry. Using the Gierer-Meinhardt prototype
model, we find the conditions in which two distinct pinning regions are formed. The first pinning type is known in the context of
variational systems while the second is structurally different due to the presence of a heteroclinic bifurcation between two uniform
states. The bifurcation also separates the parameter regions of counterpropagating fronts, leading in turn to the growth or contraction of
activator domains. These phenomena expand the range of pattern formation theory and its biomedical applications: activator domain
retraction suggests potential therapeutic strategies for patterned pathologies, such as cardiovascular calcification.
\end{abstract}

\maketitle

Reaction-diffusion (RD) systems display many kinds of pattern formation, ranging from diverse stationary patterns to breathing and/or
replicating spots and spiral waves~\cite{RD}. One important application of RD systems is to biological pattern formation and morphogenesis,
as originally suggested by Turing: reaction and diffusion of two chemical substances (``morphogens''), at different diffusion rates, can
induce the spontaneous symmetry breaking of a homogeneous state and thus form patterns~\cite{Tur:52}. Turing described a simple
activator-inhibitor system; since then, many model equations in physical, chemical, and biological contexts have been shown to form
periodic patterns via a Turing-type instability~\cite{RD,biol_morph,veg}. There is a significant difference between chemical systems and
biological media: In biology, the formation of labyrinthine patterns, for example, has been attributed to the phenomenon of
\emph{saturation}~\cite{KoMe:94, biol_morph}.

Coherent states are another important property of spatially extended media~\cite{knobloch}; we distinguish between \textit{heteroclinic}
orbits (fronts) and \textit{homoclinic} orbits (localized states) in one-dimensional (1D) physical space. In the past, the formation of
localized states in bistable systems was often linked to interactions between two front solutions~\cite{loc}. However, this approach
requires a large separation between two interacting fronts and thus is not valid for the analysis of spatial homoclinic orbits. By a
different method, it was recently shown that localized states in variational systems are intimately related to Maxwell points, and thus
nonpropagating fronts, connecting either two uniform states~\cite{KnWa:05} or a uniform and a periodic~\cite{snaking} state. In the second
case, there can be an effective broadening of the Maxwell point so that stationary fronts exist over a finite parameter
range~\cite{Pom:86}; this phenomenon is referred to as \textit{pinning}.

\begin{figure}[htb]
\includegraphics[width=3.3in]{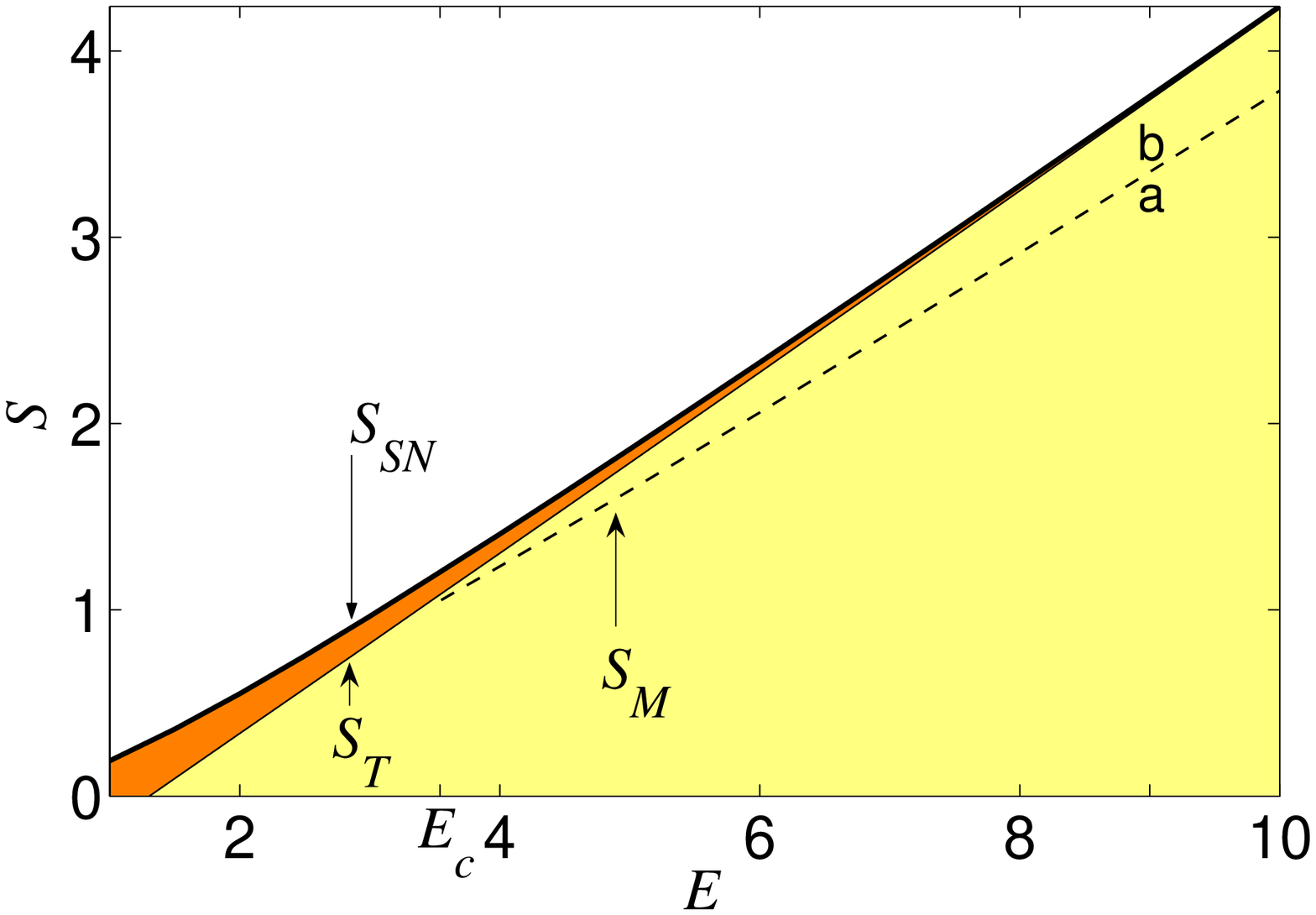}
\includegraphics[width=3.3in]{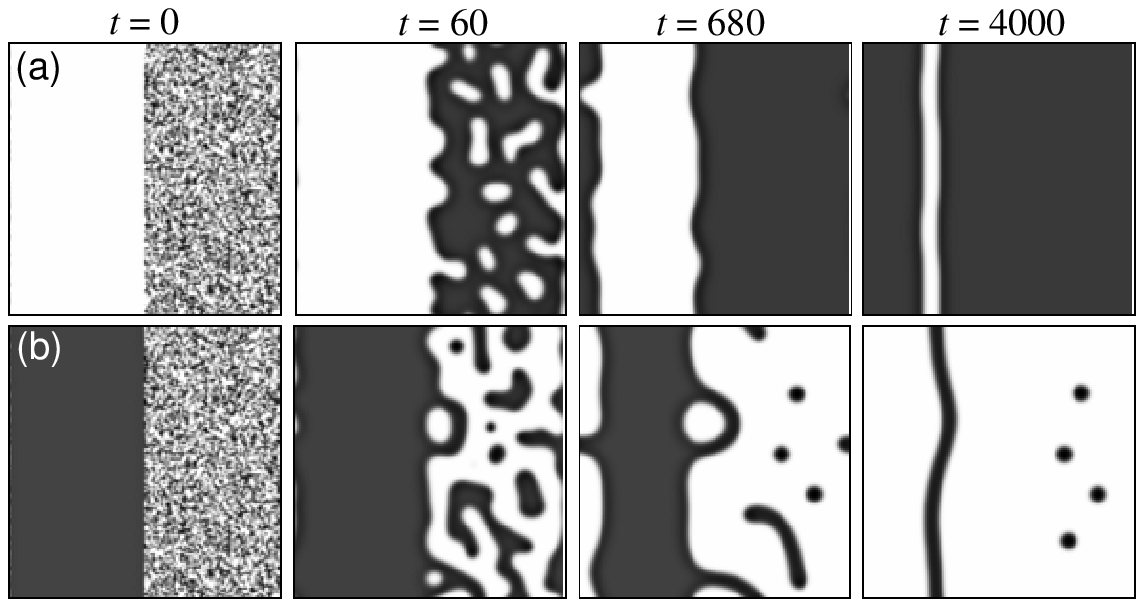}
  \caption{(Color online) Top panel: Bistability region, $S<S_{SN}$, of uniform solutions $(0,v_0)$ and $(u_+,v_+)$ in the $E-S$ plane (shaded
  region). The dark shaded region, $S_T<S<S_{SN}$, marks the linear instability of the uniform $(u_+,v_+)$ state
  to nonuniform perturbations, where $S_T$ is the onset of Turing instability. The dashed line $S=S_M$, for $E>E_c\simeq 3.54$, represents the
  stationary front between $(0,v_0)$ and $(u_+,v_+)$ states (heteroclinic bifurcation). For $S<S_M$, $(u_+,v_+)$ invades $(0,v_0)$ while for $S_M<S<S_T$
  $(0,v_0)$ invades $(u_+,v_+)$. Bottom panel: Demonstration of the activator (a) expansion and (b) contraction process at
  marked locations in the left panel, stationary front $S=S_M\simeq 3.35$ for $E=9$, via numerical integration of Eq.~(\ref{eq:GM_dimless})
  on $x=y=[0,15]$ domain with periodic boundary conditions. The frames show a gray-scale map of the $u$ field, where black denotes a high value of $u$.
  The initial condition is either uniform (a) $u=0.01$ or (b) $u=1$ in the left half of the domain, while the right half is a uniform random distribution
  between $0.01$ and $1$. Parameters: (a) $S=3.3$, (b) $S=3.4$, and for both $P=1$, $D=0.005$.}
    \label{fig:coars}
\end{figure}

While biological patterns have been extensively studied in the context of periodic patterns~\cite{biol_morph}, there is no unified theory
of coherent states (localized and fronts) and the resulting spatiotemporal dynamics in systems that involve autocatalytic
saturation~\cite{KSWW:06}. On the other hand, the effect of front asymmetry in the context of formation mechanisms of localized states is
also unclear~\cite{knobloch}. To advance these issues, we study the formation and the interaction mechanisms of coherent states, holes,
peaks and fronts, in a biological activator-inhibitor prototype model. We show that the known phenomenon of \textit{continuous branch
pinning} (CBP) can also be found in dissipative systems. In addition, we present a type of pinning, to which we refer as
\textit{discontinuous branch pinning} (DBP). In case of CBP, the existence region of holes, which is below the Turing onset, is disjunct
from the region of peaks, which is near the bistability saddle-node of the uniform states. As the control parameter is varied, the Turing
onset approaches the saddle-node, and the regions of holes and peaks overlap. This overlap results in the formation of a DBP structure of
holes via the time-independent front (hetoroclinic bifurcation), which in turn also separates regions of counterpropagating asymmetric
fronts. As a consequence, front counter-propagation and the existence of localized states give rise to an intriguing behavior: activator
domain growth or contraction accompanied by a simultaneous formation of embedded localized axisymmetric or striped states, as shown in
Fig.~\ref{fig:coars}.

We start with a phenomenological activator-inhibitor model introduced by Gierer and Meinhardt~\cite{GiMe:72}, to study biological pattern
formation. Our interest in this model stems from calcification in vascular-derived mesenchymal stem cell (VMSC) cultures~\cite{PNAS:04},
cells that are considered to be responsible for cardiovascular calcification~\cite{calc}. In particular, it is assumed from biochemistry
that the activator obeys a saturated autocatalytic reaction~\cite{satur} and promotes production of the inhibitor~\cite{inh}. The cells
also express a rapidly diffusing inhibitor~\cite{PNAS:04}. This biochemical activator-inhibitor dynamics therefore qualitatively
rationalize the framework of Gierer-Meinhardt model~\cite{GiMe:72,KoMe:94}, which in its dimensionless form reads~\cite{Yoch}
\begin{equation}\label{eq:GM_dimless}
    \begin{array}{l}
      \dfrac{\partial u}{\partial t}=D \nabla^2 u+\dfrac{u^2v^{-1}}{1+u^2}-u\,, \\
      \\
      \dfrac{\partial v}{\partial t}=\nabla^2v+Pu^2-Ev+S,
    \end{array}
\end{equation}
where $u(x,y)$ is activator, $v(x,y)$ is inhibitor, $D$ is the diffusion ratio, $P$ represents the generalized cross reaction rates ratio,
$E$ is the degradation ratio and $S$ is the generalized inhibitor source term. Due to the interest in vascular calcification~\cite{calc},
we chose here the parameters estimated in~\cite{PNAS:04} (see supporting online supplement), i.e., $P\sim\mathcal{O}(1)$,
$D\sim\mathcal{O}(10^{-2}-10^{-3})$, and focus on the two remaining parameters $S$ and $E$, which are experimentally accessible. We note
that localized peaks have also been studied in other forms of Eq. (\ref{eq:GM_dimless}), in 1D and 2D~\cite{KSWW:06,loc_gm}.

Equation~(\ref{eq:GM_dimless}) admits three uniform solutions, one of which $(u,v)=(0,v_0)\equiv (0,S/E)$ is trivial, and two nontrivial
$(u,v)=(u_\pm,v_\pm)$~\cite{coef}. Uniform $(u_-,v_-)$ states appear as unstable states from $S=0$ and annihilate with the stable
$(u_+,v_+)$ branch at a saddle-node bifurcation, $S=S_{SN}$, creating \textit{asymmetric} bistability between $(0,v_0)$ and $(u_+,v_+)$ for
$0<S<S_{SN}$. A primary requirement for both stable localized and front solutions is the linear stability of the uniform solutions to
nonuniform perturbation. Standard linear analysis of $(0,v_0)$ and $(u_+,v_+)$ yields that only the nontrivial state goes through a Turing
instability at $S=S_T$. The unstable Turing region $S_T<S<S_{SN}$ shrinks by approaching the saddle-node bifurcation, $S=S_{SN}$, as the
degradation ratio $E$ is increased (see Fig.~\ref{fig:coars}). Consequently, stable front solutions connecting the uniform $(u_+,v_+)$ and
$(0,v_0)$ states exist below $S=S_T$.

Next, we study the properties of localized states via \textit{spatial dynamics}~\cite{Cham:98} followed by numerical branch continuation
method~\cite{auto} and temporal eigenvalue numerical analysis to determine stability. We set $\partial_tu=\partial_tv=0$ and
analyze~(\ref{eq:GM_dimless}) as a reversible fourth order ordinary differential equation due to the $x\to-x$ invariance. Linearizations
about the uniform states result in steady state solutions which are proportional to $\exp(\lambda x)$, where the spatial eigenvalue
$\lambda$ satisfies a fourth order algebraic equation. For the $(u_+,v_+)$ case, three possible solutions for $\lambda$ in complex phase
space arise: (\textit{i}) for $S>S_T$ the eigenvalues split on the imaginary axis; (\textit{ii}) for $S<S_T$ they split and form a complex
quartet; and (\textit{iii}) at the Turing onset, $S=S_T$, a double multiplicity of pure imaginary eigenvalues $\lambda=\pm ik_T$ is
present, where $k_T$ is the critical Turing mode. Thus, localized states that exponentially asymptote to $(u_+,v_+)$ can be expected to
arise for $S<S_T$~\cite{snaking}. On the other hand, linearization about $(0,v_0)$ yields four real eigenvalues, $\lambda=\pm\sqrt{E}$ and
$\lambda=\pm\sqrt{D^{-1}}$. Thus, localized states that exponentially asymptote to $(0,v_0)$ are expected to arise everywhere. For finite
values of $E$, effective bifurcation of these states is at the transcritical onset, $S=0$ (the mechanism will be discussed elsewhere). In
the following, we refer to $L^+$ as ``holes'' while to $L^0$ as ``peaks'' in the background of the nontrivial and trivial activator fields,
respectively.
\begin{figure}[tp]
\includegraphics[angle=270,width=3.3in]{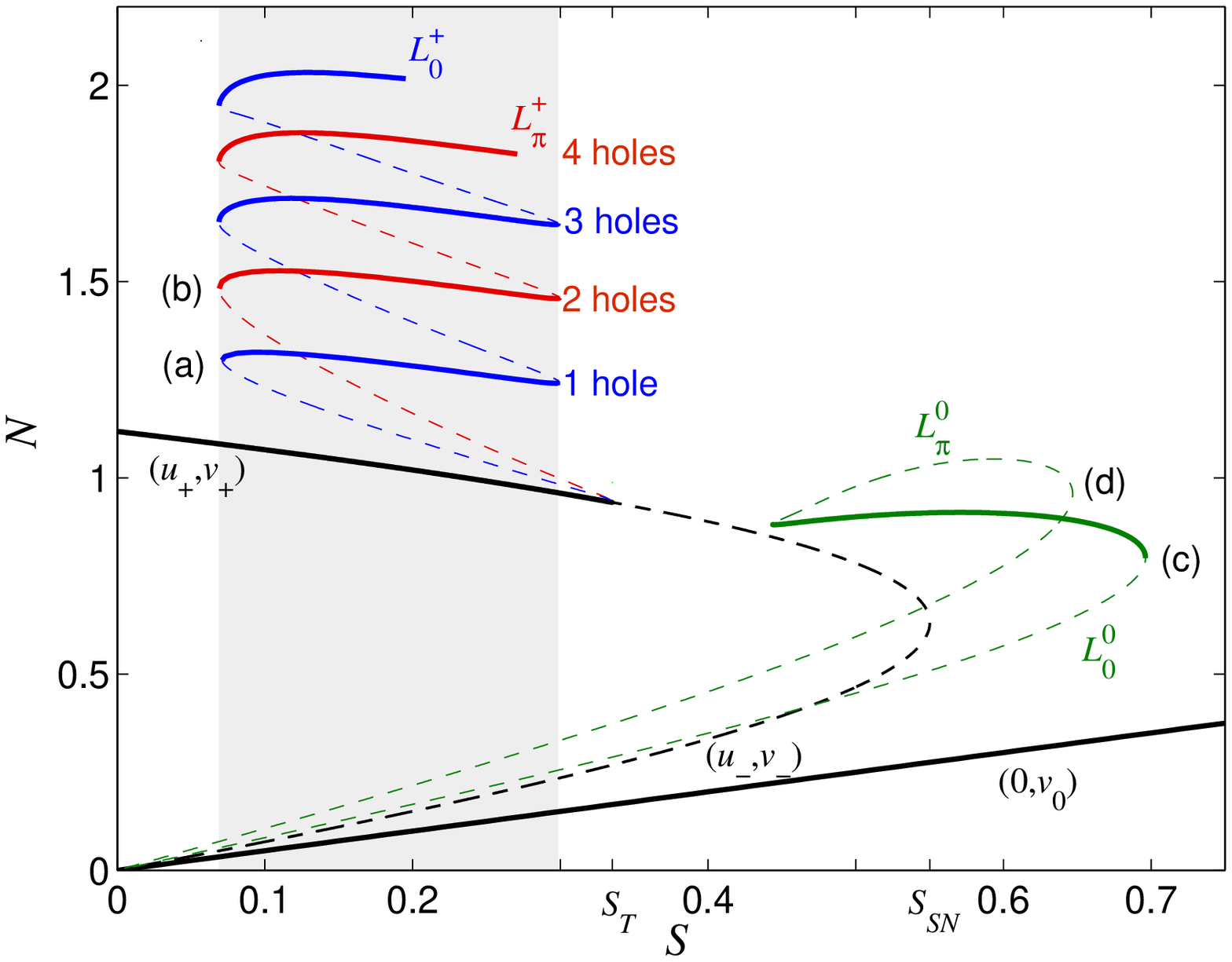}
\includegraphics[angle=270,width=3.3in]{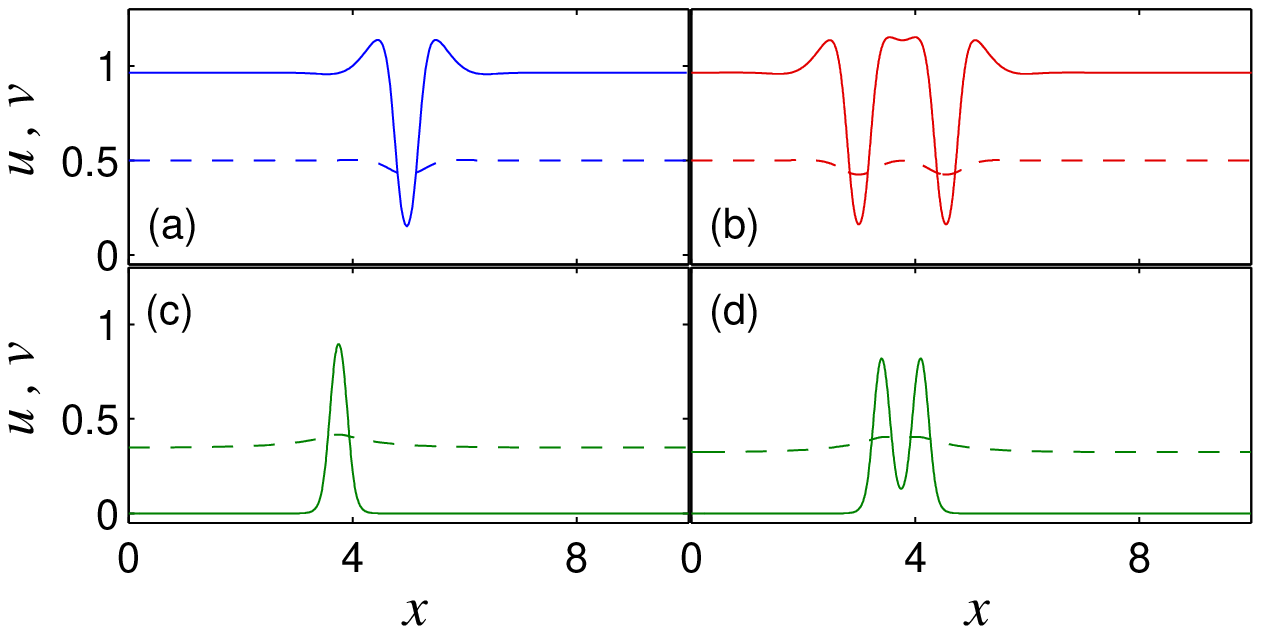}
\caption{(Color online) Top panel: Bifurcation diagram for Eq.~(\ref{eq:GM_dimless}) in 1D, showing the norm~(\ref{eq:auto}) as a function
of $S$ for branches of uniform and localized states: $L^+_0$, $L^+_\pi$ and $L^0_0$, $L^0_\pi$; solid (dashed) lines mark stable (unstable)
solutions while shaded domain represents pinning region. $S_T$ and $S_{SN}$ correspond to Turing and saddle-node onsets (see text for
details), respectively. Bottom panels: $u$ (solid) and $v$ (dashed) profiles at locations indicated in the bifurcation diagrams. The
saddle-nodes and the profiles along
  the stable portions of the branches agree with the standard time integration of (\ref{eq:GM_dimless}).
  Parameters: (a,b) $S\simeq 0.071$ (c) $S\simeq 0.696$, (d) $S\simeq 0.647$ and for all $E=2$, $D=0.005$, $P=1$.} \label{fig:bif_profE2}
\end{figure}
\begin{figure}[tp]
\includegraphics[angle=270,width=3.3in]{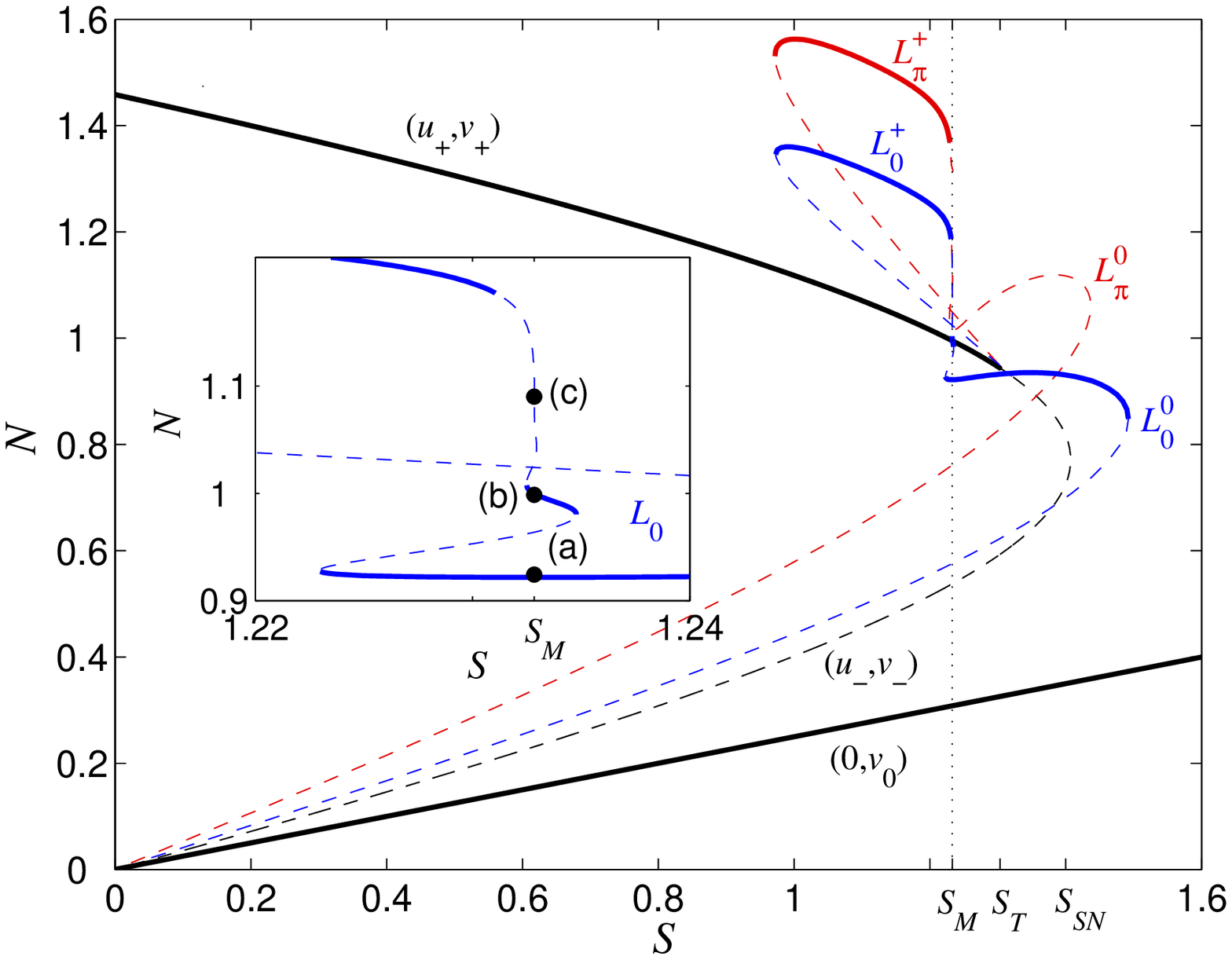}
\begin{minipage}[t]{0.6\linewidth}
\includegraphics[angle=270,width=1.8in]{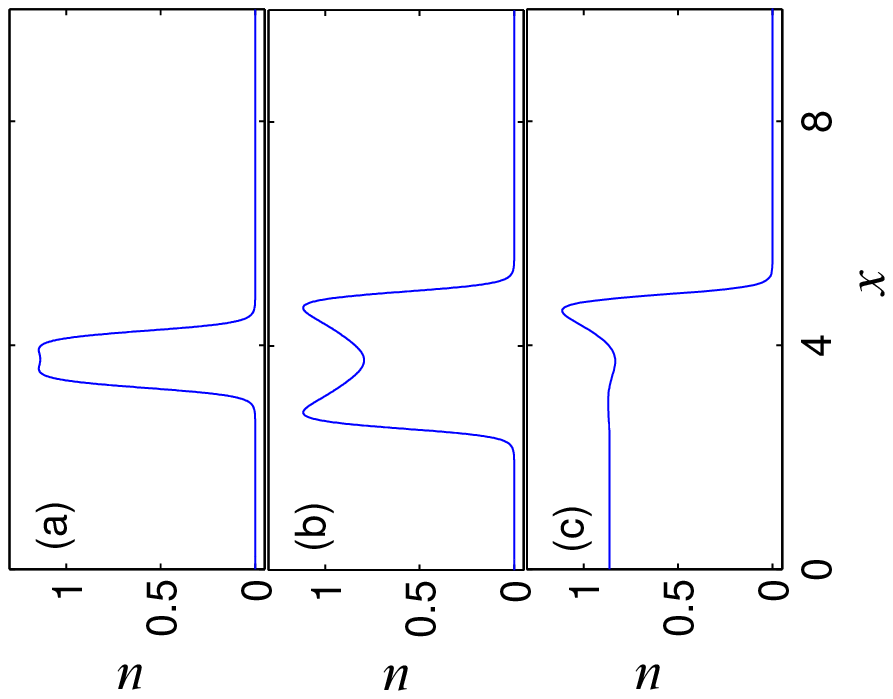}
\end{minipage}\hfill
\begin{minipage}[t]{0.4\linewidth}
  \caption{(Color online) Top panel: Bifurcation diagram for $E=4$ (solid (dashed) lines mark stable (unstable) solutions). The inset in the top panel clarifies the behavior of
  $L^+_0$ and $L^0_0$ branches at $S=S_M\simeq 1.233$. Bottom panels: Profiles of $u$ at
  locations indicated in the inset. Note the asymmetry of the heteroclinic orbit (half of the cycle) in (c).
  Other parameters as in Fig.~\ref{fig:bif_profE2}.} \label{fig:bif_profE4}
\end{minipage}
\end{figure}

To study the localized solutions \textit{far} from the onset, we implement a numerical branch continuation method~\cite{auto}; the results
are presented as a function of $S$ in terms of the norm
\begin{equation}\label{eq:auto}
    N=\sqrt{\frac{1}{L}\int_0^L \Bra{u^2+v^2+\bra{\partial_x u}^2+\bra{\partial_x v}^2} \mathrm{d}x}\,,
\end{equation}
where $L$ is the spatial period; the domain size used in our numerical calculations is much larger than the spatial period $2\pi/k_T$. We
distinguish between two qualitatively distinct regions (see Fig.~\ref{fig:coars}): $E<E_c$, where the holes and peaks do not overlap, and
$E>E_c$ (see Fig.~\ref{fig:bif_profE2}), where peaks and holes regions do overlap, and reconnect at $S=S_M$ via the stationary heteroclinic
cycle between $(u_+,v_+)$ and $(0,v_0)$ (see Fig.~\ref{fig:bif_profE4}).

First, we discuss the $E<E_c$ case. Localized hole solutions, $L^+$, and periodic Turing states (not shown) bifurcate subcritically as
unstable small amplitude states~\cite{Yoch}, from the Turing onset, $S=S_T$, and form a CBP region (shaded region in
Fig.~\ref{fig:bif_profE2}), as also happens in variational systems~\cite{snaking}. Since the localized holes bifurcate from a nontrivial
state, only two even parity solutions with phase shifts of $\pi$ can form. Thus, we termed a branch with an odd number of holes as $L^+_0$
[Fig.~\ref{fig:bif_profE2}(a)] and the one with an even number as $L^+_\pi$ [Fig.~\ref{fig:bif_profE2}(b)]. Each stable branch of odd and
even hole solutions (inside the shaded region) indicate increasing (with $N$) number of holes~\cite{snaking}. Here we used a relatively
small domain which limits further excursions of the two branches (Fig.~\ref{fig:bif_profE2}), while on an infinite domain, the branches
contain an infinite number of saddle-node points. On the other hand, the localized states that are homoclinic to the $(0,v_0)$ state
effectively bifurcate from $S=0$ and do not form a pinning region, due to real eigenvalues. These states contain only a single stable
branch of $L^0_0$ while the other even branch $L^0_\pi$ is unstable [see profiles (c,d) in Fig.~\ref{fig:bif_profE2}], denoting a repulsive
interaction between two neighboring peaks. Here the subscripts $0$ and $\pi$, are artificial notations.

Next, we turn to a discussion of the second case, $E>E_c$. We showed in Fig.~\ref{fig:coars}, that an increase in $E$ results in $S_T \to
S_{SN}$. Above $E= E_c\simeq 3.54$, this shift causes an overlapping between the CBP region of hole states and the peak states that are
present near the saddle-node $S=S_{SN}$. This hole-peak interaction allows formation of a time-independent asymmetric front (via
heteroclinic bifurcation) at $S=S_M$, connecting the trivial and the nontrivial states (see Fig.~\ref{fig:bif_profE4}). All the localized
state branches that were present in the $E<E_c$ case, now effectively collapse nonmonotonically at $S=S_M$. The single and the double hole
state, $L=L^+$, still bifurcate from the Turing onset (see Fig.~\ref{fig:bif_profE4}), but all other branches (not shown here for
simplicity) arise from and reconnect after a single excursion, back to $S=S_M$. This DBP region is structurally different from the
classical CBP, described for $E<E_c$ and in~\cite{snaking}. The $L^0$ branches also go through a significant change: while $L_\pi^0$
remains unstable, the $L_0^0$ branch exhibits decreasing oscillations around $S=S_M$ (see inset of Fig.~\ref{fig:bif_profE4}). This
behavior is known to give rise to additional localized states~\cite{KnWa:05,YBK:06}, with weak spatial oscillations around the primary peak
[see Fig.~\ref{fig:bif_profE4}(b)].

The activator expansion or contraction and formation of isolated stripes and axisymmetric localized states (Fig.~\ref{fig:coars}) is now
transparent. The effective merging point, $S=S_M$, separates regions of counterpropagating single fronts. Thus, opposite front velocities
and the simultaneous presence of stable localized holes, peaks and stripes, result in a nontrivial behavior for bistable nonvariational
systems with a broken parity symmetry of fronts [see Fig.~\ref{fig:bif_profE4}(c)]. In the absence of localized states, the system
converges to a finite uniform activator state, a phenomenon that is qualitatively similar to coarsening in variational systems~\cite{loc}.
An additional important aspect is the stability of coherent structures to curvature and transverse perturbations in 2D. While stability
thresholds of localized states may indeed change in 2D, the 1D analysis delimits the regions where stable localized states can also be
obtained in 2D~\cite{YBK:06,Yoch}. By contrast, instability to transverse perturbations of the front solutions or large circular domains
may lead to the formation of periodic patterns that resemble periodic patterns arising from the Turing instability~\cite{trans}. In our
parameter range these fronts are transversally stable, but we cannot exclude the existence of transverse instabilities in general.

We have demonstrated a mechanism for the expansion and contraction of activator domains, and the respective formation of localized states,
in a bistable activator-inhibitor system with saturated autocatalysis (Fig.~\ref{fig:coars}). This mechanism stems from the formation of an
asymmetric heteroclinic orbit (at a critical parameter value), merging the branches of both hole and peak states. Turing instabilities are
found to be important for such behavior, and also for a discontinuous branch pinning structure, as shown in Fig.~\ref{fig:bif_profE4}.
Since we used global bifurcation methods, we expect similar scenarios to arise in other bistable systems as well, precisely for the same
reason that pinning~\cite{snaking} is found in variational models and in our dissipative system. Furthermore, our theoretical results
suggest a distinct spatiotemporal behavior in biological systems exhibiting saturation~\cite{KoMe:94}. The concept of activator contraction
has potential biomedical applications, for example, localized activator-induced ``nodules'' are common in biomedicine, such as spotty
formations in atherosclerotic calcification~\cite{ATD:04}. The ability to produce conditions of ``activator contraction'' to reduce or
eliminate these localized spots, therefore has intriguing therapeutic possibilities.
\\

We thank L.L. Demer, Y. Tintut, K. Bostr\"{o}m, E. Knobloch, and J. Burke for stimulating discussions.

\end{document}